\documentclass[
preprint,
 amsmath,amssymb,
 aps, physrev,
]{revtex4-2}

\usepackage{graphicx}
\usepackage{dcolumn}
\usepackage{bm}
\usepackage{multirow}
\usepackage{adjustbox,upgreek}
\usepackage[mathlines]{lineno}

\nolinenumbers 

\usepackage{hyperref}
\hypersetup{
 colorlinks=true,
 linkcolor=blue,
 citecolor=blue}
 
\begin{document}

\preprint{preprint}

\title{Gaussian splatting holography}%

\author{Shuhe Zhang}
\author{Liangcai Cao}%
 \email{Contact author: clc@tsinghua.edu.cn}
\affiliation{%
 Department of Precision Instruments, Tsinghua University, Beijing 100084, China
}%

\date{\today}

\begin{abstract}
In-line holography offers high space–bandwidth product imaging with a simplified lens-free optical system. However, in-line holographic reconstruction is troubled by twin images arising from the Hermitian symmetry of complex fields. Twin images disturb the reconstruction in solving the ill-posed phase retrieval. The known parameters are less than the unknown parameters, causing phase ambiguities. State-of-the-art deep-learning or non-learning methods face challenges in balancing data fidelity with twin-image disturbance. We propose the Gaussian splatting holography (GSH) for twin-image-suppressed holographic reconstruction. GSH uses Gaussian splatting for optical field representation and compresses the number of unknown parameters by a maximum of 15 folds, transforming the original ill-posed phase retrieval into a well-posed one with reduced phase ambiguities. Additionally, the Gaussian splatting tends to form sharp patterns rather than those with noisy twin-image backgrounds as each Gaussian has a spatially slow-varying profile. Experiments show that GSH achieves constraint-free recovery for in-line holography with accuracy comparable to state-of-the-art constraint-based methods, with an average peak signal-to-noise ratio equal to 26 dB, and structure similarity equal to 0.8. Combined with total variation, GSH can be further improved, obtaining a peak signal-to-noise ratio of 31 dB, and a high compression ability of up to 15 folds. 
\end{abstract}

\maketitle
Accurately detecting wave propagation in both space and time is crucial for studying wave–matter interactions. Holography is an advanced imaging technique that records both the amplitude and phase of light waves by capturing interference patterns \cite{rosen2024roadmap}. Holography allows for the recording of the optical wavefront, making holography useful in fields like quantitative phase imaging \cite{mann2005high,huang2024quantitative}, 3D visualization \cite{kim2014white,shi2021towards}, and non-invasive measurements \cite{paturzo2018digital}. Among the various types of holography, in-line holography offers benefits such as high spatial bandwidth usage, single-shot imaging, and simpler optical setups \cite{Gabor1948}. However, in-line holography has a major challenge: twin-image artifacts, which can significantly degrade the quality of phase retrieval \cite{PhysRevLett.98.233901, PhysRevLett.121.093902,wu2020single,gao2023iterative}. Twin images arise from the phase-conjugate symmetry of interference patterns, leading to unwanted artifacts. Mathematically, the twin-image arises from the ill-posed nature of phase retrieval for in-line holography from single intensity-only measurement, meaning the number of unknown parameters (the complex amplitude of the sample) exceeds the number of known parameters, leading to phase ambiguity \cite{popescu2021large}. For instance, a hologram with a size of $1k\times 1k$ pixels has $1k\times1k\times2$ unknown parameters to recover. Using pure intensity/phase prior knowledge can reduce the unknowns by half. But twin-image removal is still challenging.

Current solutions to the twin-image issue originate from a single ideal: introducing more known parameters than the unknowns to make the phase retrieval well-pose. Interference holography introduces the reference beam to suppress the phase ambiguity \cite{kim2014white}. Phase diversity methods introduce more intensity measurement under different defocus, positions \cite{paxman1992joint,rodenburg2004phase}, or illuminations \cite{jiang2023spatial} to suppress phase ambiguity. Algorithmic methods such as regularization \cite{Brady:09, PhysRevLett.121.093902} or deep learning \cite{rivenson2019deep,wang2020phase,chang2021large} introduce image priors explicitly or implicitly to suppress the phase ambiguity. The aforementioned methods either suffer from complex image acquisition strategies or struggle with tuning the parameters to balance data fidelity and reconstruction quality.

Instead of introducing additional known parameters to suppress phase ambiguities, compressing the number of unknown parameters can be a potential solution to address the ambiguities that have not yet been studied. From parameters compressing, the Gaussian splatting is a state-of-the-art image compressor that enables super-fast compression of a given image or a 3D scene by folds, while maintaining high qualities \cite{KKLD23, GaussianImage,niedermayr2024compressed,chen2024hac}. For example, a single 2D Gaussian consists of 7 unknown parameters, and an image of $1k \times 1k$ pixels can be represented by 9000-20000 Gaussians with a peak signal-to-noise ratio (PSNR) $\ge 40$ db (\textbf{Supplementary Note 1}). The parameters are compressed by 7 to 16 folds \cite{GaussianImage}. 

Inspired by Gaussian splatting, this letter presents a regularization-free computational framework for in-line holography, termed Gaussian splatting holography (GSH), by introducing Gaussian splatting for compressive image representations. The GSH utilizes the Gaussian splatting to render the complex amplitude of the sample for parameter compression \cite{GaussianImage}. With the compressed Gaussian representation, the quantity of unknown parameters dramatically decreases and the inverse problem becomes well-posed. Accordingly, we hypothesize that the GSH can achieve twin-image-free in-line holography without regularization. Simulation and experimental studies are presented later in this letter to support our hypothesis. 

\begin{figure}
    \centering
    \includegraphics[width=0.75\linewidth,trim = 0 0 395 0,clip]{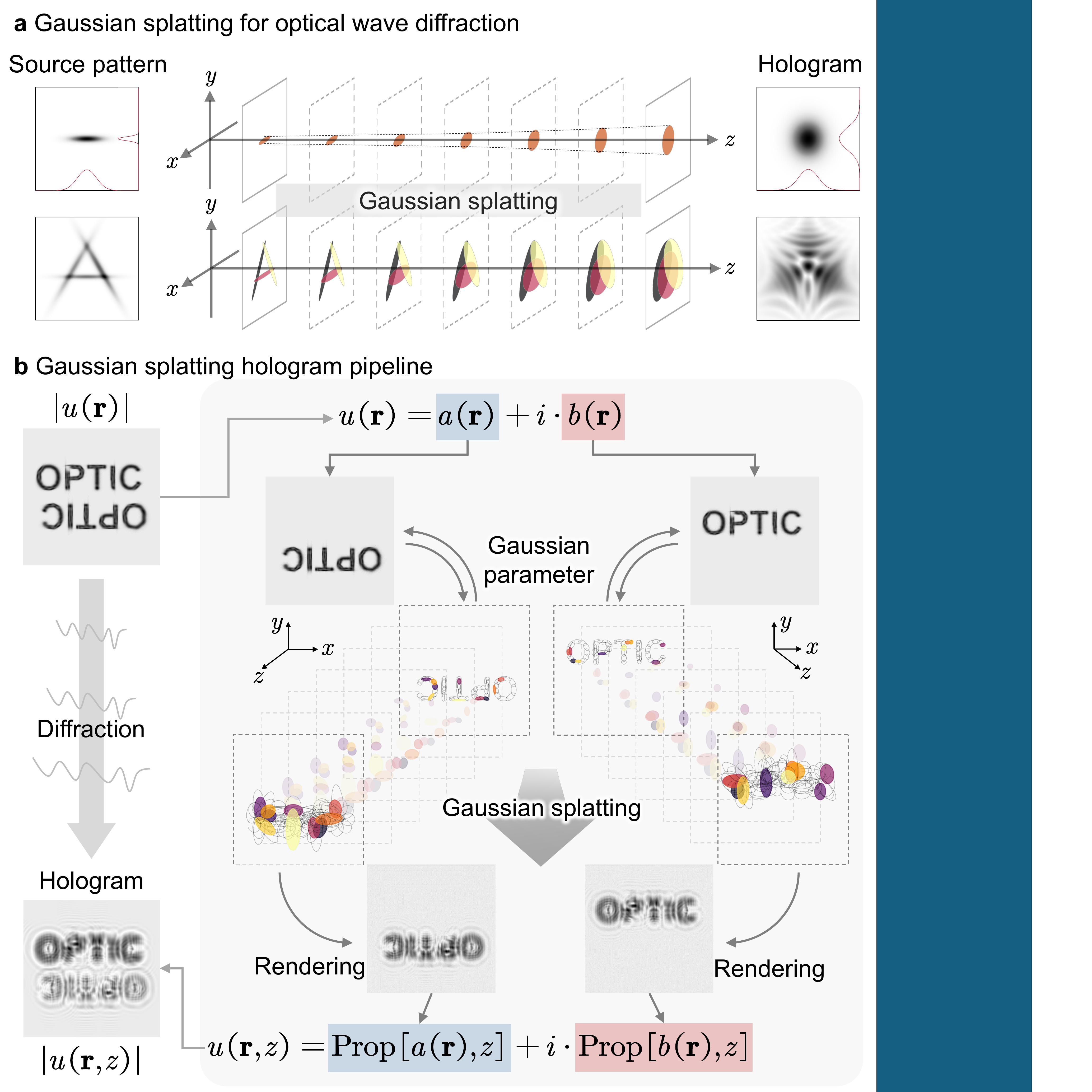}
    \caption{Sketch of Gaussian splatting for in-line holography formation. (\textbf{a}) Gaussian splatting for a single Gaussian and 3 Gaussians. (\textbf{b}) Gaussian splatting for general in-line holography representation. The real and imaginary parts of a sample optical wave are separately expressed as the superposition of a series of independent-propagating Gaussians. The final observed hologram is the coherent summation and intensity-only detection of all propagated Gaussians.}
    \label{fig: fig1}
\end{figure}

Figure \ref{fig: fig1} (\textbf{a}) shows the concept of the GSH. Any image can be expressed by a group of Gaussians, with different positions, sizes, orientations, and amplitude values. Propagation properties of a single Gaussian are similar to an elliptical Gaussian beam \cite{carter1972electromagnetic,cornolti1990elliptic,heyman2001gaussian,cai2002decentered}, shown in the first row of Fig. \ref{fig: fig1} (\textbf{a}), where the orange ellipses draw the contour of the waist of the Gaussian beam at given $z$-planes. In the GSH, the hologram of a single Gaussian after splatting denotes the diffraction pattern of the single Gaussian after propagation. Intricate image patterns can be rendered by introducing more Gaussians with different shapes. The second row in Fig. \ref{fig: fig1} (\textbf{a}) shows a pattern of the letter "A" composed of 3 Gaussians and their propagation properties. After splatting to the target plane, the observed hologram is the intensity component of the interference pattern of all 3 Gaussians. 

Figure \ref{fig: fig1} (\textbf{b}) delineates the overall forward pipeline of our GSH for single-shot, in-line holography. The sample's complex amplitude $u(\mathbf{r})$ is expressed as the summation of its real part $\Re [u(\mathbf{r})] = a(\mathbf{r}) \in \mathbb{R} $, and its imaginary part $\Im [u(\mathbf{r})] = b(\mathbf{r}) \in \mathbb{R}$, $\mathbf{r} = (x,y)$ denoting the spatial coordinates. With Gaussian splatting, the real part, for example, is further expressed by the superposition of a series of Gaussians, given as \cite{KKLD23, GaussianImage}
\begin{equation}
    a(\mathbf{r}) = \sum_{l=1}^L v_l\cdot \sigma_l\cdot G_l(\mathbf{r}), \label{Eq. 1}
\end{equation}
where $v_l$ and $\sigma_l$ denote the strength and transparency of the Gaussian, respectively. $G(x,y) \in \mathbb{R}$ denotes the pattern of a Gaussian, given as 
\begin{equation}
    G_l(\mathbf{r}) = \exp\left[-\frac{1}{2}(\mathbf{r}-\mathbf{c}_{l})^\top\mathbf{\Sigma}^{-1}_l(\mathbf{r}-\mathbf{c}_l)\right], \label{Eq. 2}
\end{equation}
where $\mathbf{c}_l=(x_{c,l},y_{c,l})$ denotes the central of the $l$-th Gaussian, $\mathbf{\Sigma} = \mathbf{RS}\mathbf{S}^\top\mathbf{R}^\top$ is the covariance of the Gaussian, where $\mathbf{R}_l$ is 2D rotation matrix and $\mathbf{S}_l$ is 2D scaling matrix. 

A similar Gaussian superposition applies to the imaginary part $b(\mathbf{r})$, and thus two sets of Gaussians are used to represent the $a$ and $b$, respectively. Gaussians for the real and imaginary parts of propagate independently in a non-paraxial manner \cite{dickson1970characteristics,agrawal1979gaussian,alda2003laser,zhang2018skew}.
According to Eqs. (\ref{Eq. 1}-\ref{Eq. 2}), each Gaussian has 7 unknown parameters, so the real part $a(\mathbf{r})$ has a total of $7L$ unknown parameters to be solved. For both real and imaginary parts, a total of $14L$ unknown parameters are to be solved. Please refer to \textbf{Supplementary Note 2} for details.


With the captured hologram $I(\mathbf{r},z)$, and propagation distance $z$, the unknown $14L$ Gaussian parameters can be learned by minimizing the energy function provided as 
\begin{equation}
\mathcal{L}=\iint_\mathbf{r}   \left (   \sqrt{I(\mathbf{r},z)}  - \left | \mathcal{F}^{-1}\left \{ \mathcal{F}\left [ u(\mathbf{r}) \right ]\exp\left ( ik_zz \right ) \right \} \right | \right )^2\text{d}^2\mathbf{r}. \label{Eq. 4}
\end{equation} 
Here, the angular spectrum diffraction describes wave propagation in in-line holography \cite{goodman2005introduction}, where $\mathcal{F}$ denotes the Fourier transform. $k_z = \sqrt{k^2-4\pi^2( f_x^2+f_y^2)}$. $k$ is the wave number. $f_x$, and $f_y$ are the spatial frequency in $x$- and $y$-direction. The diffraction is independently applied to the real and imaginary parts of the target wave due to the linearity of the wave equation. The measured hologram is the intensity of the interference of the complex amplitude of the real and imaginary parts after diffraction.  

Eq. (\ref{Eq. 1}) is the main equation of Gaussian splatting. From an image-processing perspective, $a$ represents an image that is synthesized by combining a series of Gaussians $G$ with different widths, orientations, and heights, as described in Eq. (\ref{Eq. 2}) \cite{GaussianImage}. A group of Gaussians can fit the image by training the parameters based on the difference between the synthetic image and the ground truth \cite{KKLD23}, by minimizing the energy function in Eq. (\ref{Eq. 4}).

\begin{figure}
    \centering
    \includegraphics[width=0.75\linewidth,trim = 210 704 250 0,clip]{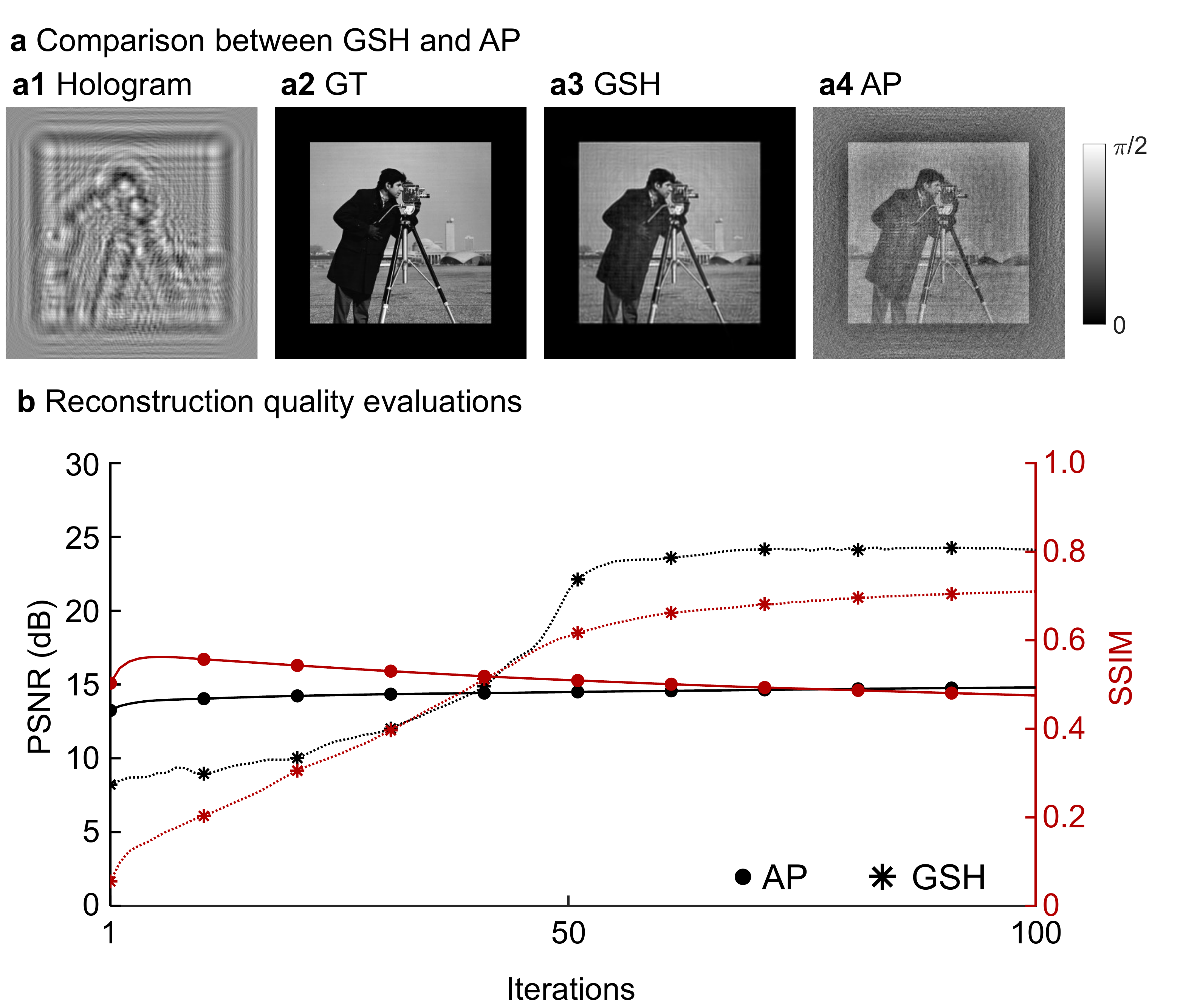}
    \caption{Simulation study of GSH on pure-phase sample. (\textbf{a}) Comparison between GSH and AP. \textbf{a1} Simulated hologram. \textbf{a2} Ground truth of phase pattern. \textbf{a3} Reconstruction results of GSH. \textbf{a4} Reconstruction results of alternative projection (AP). (\textbf{b}) Evolution of image quality matrices.}
    \label{fig: fig3}
\end{figure}

The GSH was tested first through a simulation with a pure-phase target (see Fig. \ref{fig: fig3}) with $256 \times 256$ pixels. 3600 Gaussians were used to represent the phase pattern. Other simulation parameters are provided in \textbf{Supplementary Note 3} for details. The image is padded with 50 pixels on all sides. The diffraction pattern of the target is shown in Fig. \ref{fig: fig3} (\textbf{a1}), and the ground truth phase pattern is shown in Fig. \ref{fig: fig3} (\textbf{a2}). The pixel values are normalized between 0 and 0.5$\pi$.  Each pixel represents 2.56 $\upmu$m, with an illumination wavelength of 0.532 $\upmu$m and a diffraction distance of 5000 $\upmu$m. Conventional phase retrieval requires solving for 126,736 parameters, while 3600 Gaussians were used to represent the phase pattern, and the GSH solved only 25,200 parameters, reducing the number of unknowns by 5-fold. The Lion optimizer \cite{chen2024symbolic} was used to train the Gaussian, and the learning rate is first set to 0.004, and decayed by 0.5 for every 400 iterations. The Gaussian is trained for a total of 5000 iterations. With the enhancement of the Gsplat package \cite{ye2024gsplat}, the total training time is 8s with an NVIDIA RTX 3090 graphic card. 

\begin{figure}
    \centering
    \includegraphics[width=0.75\linewidth,trim = 17 260 17 0,clip]{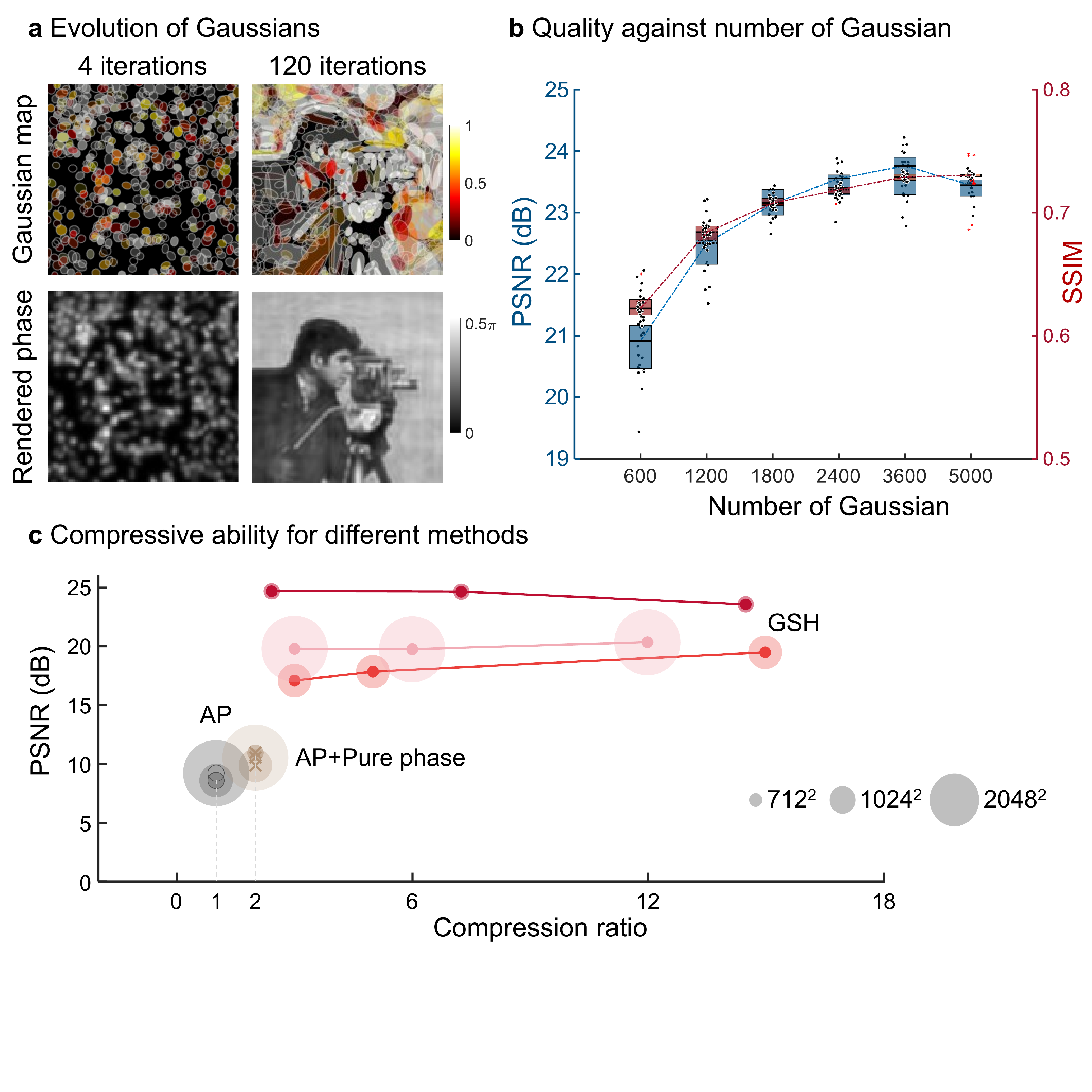}
    \caption{Quality of GSH w.r.t number of Gaussians. (\textbf{a}) Gaussian map at 4, and 120 iterations. (\textbf{b}) Image quality matrices for results reconstructed by different numbers of Gaussians. (\textbf{c}) Compression ratio on different hologram sizes for the alternative projection and GSH, without/with regularization.}
    \label{fig: Map}
\end{figure}

The results of GSH shown in Fig. \ref{fig: fig3} (\textbf{a3}) are compared with those of the conventional alternative projection (AP) algorithm \cite{Elser:03,jaganathan2016phase}, as shown in Fig. \ref{fig: fig3} (\textbf{a4}), which uses only a pure-phase constraint. The AP result is affected by twin-image artifacts, which appear as noise in the background (Fig. \ref{fig: fig3} (\textbf{a4})), whereas the GSH shows a clear background (Fig. \ref{fig: fig3} (\textbf{a3})). The evolutions of the loss function, PSNR, and structural similarity index (SSIM) are shown in Fig. \ref{fig: fig3} (\textbf{b}), measuring the quality of reconstruction \cite{hore2010image}. Interestingly, although the AP algorithm achieves a lower loss function, indicating a better fit to the measured hologram, its PSNR and SSIM scores are worse than the GSH due to the twin-image artifacts.
As the Gaussians are trained iteratively, we can monitor how they gradually fit the phase pattern. Fig. \ref{fig: Map} (\textbf{a}) shows the Gaussian map and the corresponding rendered phase pattern at 4 and 120 iterations. The color map indicates the weight of each Gaussian (denoted as $v_l$), while the transparency represents the value of $\sigma_l$. In the early iterations, the Gaussians are randomly placed, as shown in Fig. \ref{fig: Map} (\textbf{a}), and the synthetic image is a mix of randomly distributed Gaussians. As training progresses, the Gaussians shift and reshape, gradually forming key image features, such as the face of the cameraman. Over time, the synthetic image becomes more similar to the ground truth.

The quantity of used Gaussians is a key hyperparameter. In this simulation, we observed that the PSNR increases as the quantity of Gaussians used in the reconstruction increases, as shown in Fig. \ref{fig: Map} (\textbf{b}). After reaching around 2400 Gaussians, the PSNR plateaus and slightly decreases or fluctuates after 3600 and 5000 Gaussians. Similar to PSNR, the SSIM improves as the quantity of Gaussians increases, showing steady growth in Fig. \ref{fig: Map} (\textbf{b}). SSIM also stabilizes after around 2400 Gaussians, achieving near-saturation levels. The improvements in PSNR and SSIM w.r.t. the number of Gaussians imply that the GSH effectively captures both the pixel-level (PSNR) and structural (SSIM) information of the phase pattern as the number of Gaussians increases. 

A higher quantity of Gaussians improves the reconstruction quality as the model has more degrees of freedom to fit the phase pattern. However, beyond a certain threshold, relying upon the size of the image, adding more Gaussians provides diminishing returns or may even slightly degrade performance. The degradation could be due to overfitting or numerical instability introduced by too many parameters.  
It is suggested that the number of Gaussians should be determined based on the size of the input hologram, where $ (MN/70) \le L\le(MN/7)$, where $M$ and $N$ is the width and the height of the hologram. We further introduce adaptive Gaussian densification to adaptively control the quantity of Gaussians so that the optical field can be represented by a sufficient number of Gaussians to maintain fine structures while not wasting unnecessary computational resources. Please refer to \textbf{Supplementary Note 4} for details.

The compression ratios, calculated by the ratio between the number of known and unknown parameters, for the AP and the GSH are plotted in Fig. \ref{fig: Map} (\textbf{c}), with reconstruction PNSR. The GSH has a high compression ratio between 4 to 15, and obtains a significant improvement in PSNR compared to AP and AP with total variation regularization (Reg-AP), even without regularization. The GSH can be combined with regularization to improve its reconstruction quality. As plotted by the blue bubble, the GSH with total variation (TV) \cite{rudin1992nonlinear,rudin1994total,chambolle2004algorithm} offers high-quality reconstruction in terms of PSNR, particularly at high compression ratios. The inclusion of TV regularization allows it to achieve PSNR values above 30 dB, maintaining strong reconstruction quality even at higher compression levels.

\begin{figure*}
    \centering
    \includegraphics[width=0.96\linewidth,trim = 0 10 800 0,clip]{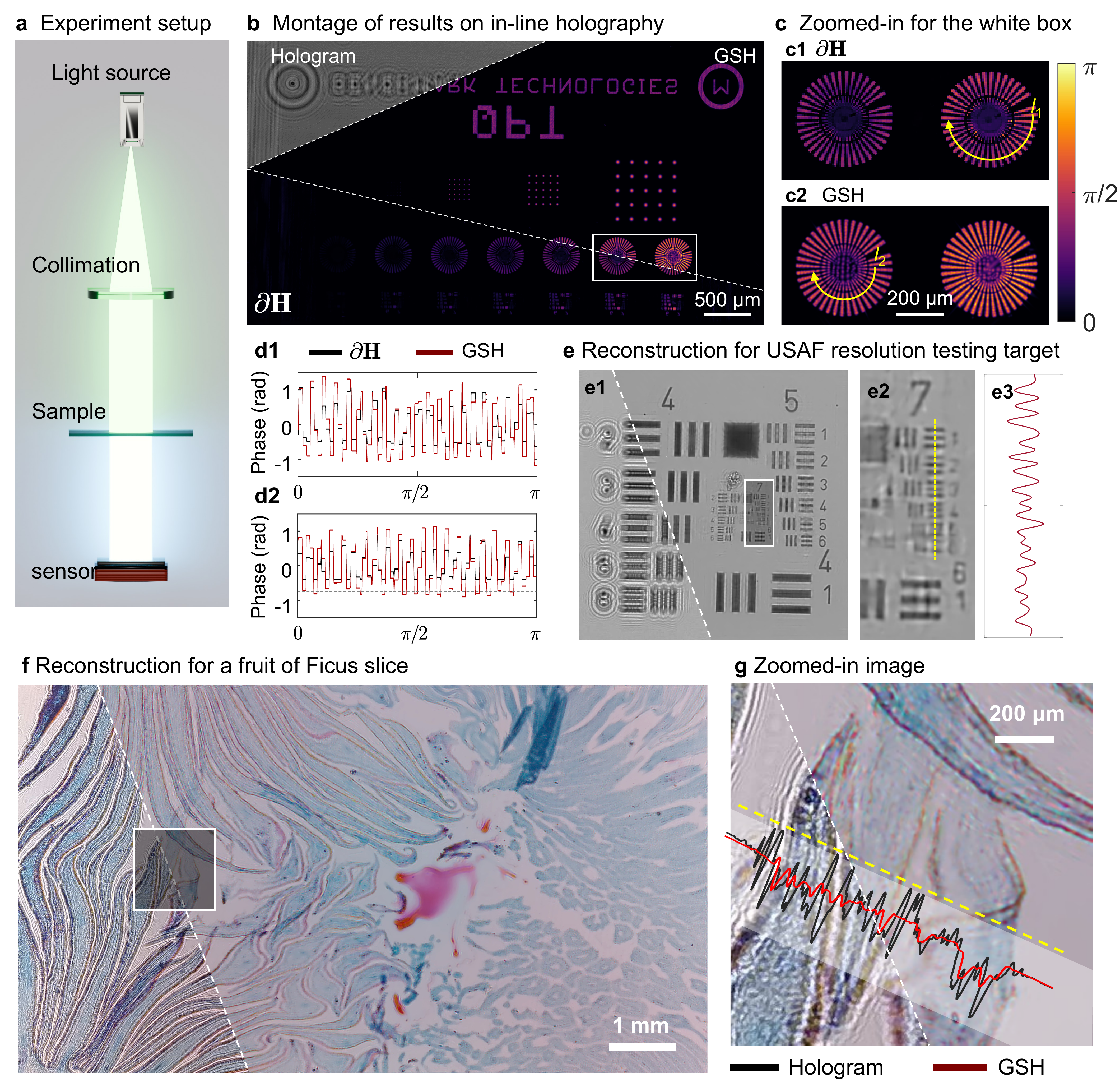}
    \caption{Experimental studies for GSH. (\textbf{a}) Experimental layout. (\textbf{b}) Montage of the hologram, results from $\partial \mathbf{H}$, and proposed GSH methods for a quantitative phase target. (\textbf{c}) Zoomed-in image for the area in the white box in (\textbf{b}) reconstructed with (\textbf{c1}) $\partial \mathbf{H}$, and (\textbf{c2}) GSH. (\textbf{d1}) and (\textbf{d2}) Phase profile along the curve in $l_1$ and $l_2$. (\textbf{e}) Reconstruction of a USAF resolution testing target using GSH. \textbf{e1} Montage of the hologram, and results from GSH. \textbf{e2} Zoomed-in image for the gray box in (\textbf{e1}). (\textbf{e3}) Amplitude profile along the yellow line. (\textbf{f}) Reconstruction of a Ficus slice. (\textbf{g}) Zoomed-in image for the gray box in \textbf{f}.}
    \label{fig: fig4}
\end{figure*}

The GSH was further evaluated experimentally shown in Fig. \ref{fig: fig4} (\textbf{a}) using a quantitative phase target, a USAF resolution testing target, and a sliced sample. (see Fig. \ref{Eq. 4}). The hologram shown in Fig. \ref{Eq. 4} (\textbf{b}) has a size of $945 \times 517$ pixels, and 53,000 Gaussians were used for the reconstruction. This reduced the number of unknown parameters by a factor of 1.31. The camera sensor pixel size is 5.86 $\upmu$m, and the diffraction distance is 8780 $\upmu$m, with an illumination wavelength of 0.660 $\upmu$m. Fig. \ref{Eq. 4} (\textbf{b}) shows the phase pattern reconstructed using the GSH with no regularization, relying only on the loss function in Eq. (\ref{Eq. 4}). We compared this with the state-of-the-art differential holography ($\partial \mathbf{H}$) \cite{chen2023h,chen2025differentiable}. GSH outperforms $\partial \mathbf{H}$, showing clearer and more detailed phase patterns, especially finer and high-frequency features. In Fig. \ref{fig: fig4}, zoomed-in images (in blue boxes) highlight the sharper phase pattern from GSH compared to $\partial \mathbf{H}$, with Fig. \ref{fig: fig4} (\textbf{d1}) and Fig. \ref{fig: fig4} (\textbf{d2}) showing the quantitative phase profiles. The contrast in GSH is higher, demonstrating its advantage in capturing both low- and high-frequency phase components.

The resolution of Gaussian is further validated on the USAF resolution testing target. Here the pixel size of the camera is 2.7 $\upmu$m. The captured diffraction pattern of the USAF and the reconstructed amplitude are shown in Fig. \ref{Eq. 4} (\textbf{e1}). From the zoomed-in image in Fig. \ref{Eq. 4} (\textbf{e2}) and the amplitude profile plotted in Fig. \ref{Eq. 4} (\textbf{e3}), the line structure of the group 7 element 4 can be resolved, denoting a spatial resolution of 181 lp/mm. The width of the stripe for group 7 element 4 is 2.76 $\upmu$m, which approaches the pixel pitch. 

Finally, we performed the lens-free whole-slide microscopy of a fruit slice with $3840 \times 2160$ pixels. A total of 450,000 Gaussians were used.  Fig. \ref{Eq. 4} (\textbf{f}) shows the reconstruction of a thin slice of a Ficus plant sample using GSH. The image provides a large-scale view of the structural details of the Ficus slice, with distinct textures and patterns observable in the reconstructed image. Detailed structures can be observed from the zoomed-in image in Fig. \ref{Eq. 4} (\textbf{g}), highlighting the effectiveness of GSH in suppressing twin-image and providing reliable signal reconstructions. \textbf{Supplementary Notes 5} and \textbf{Note 6} provide additional experimental studies of GSH.

In conclusion, the GSH is a new image compression and reconstruction method for holographic imaging, making the inverse problem well-posed without regularization. The compressive parameters feature enables accurate reconstruction of both amplitude and phase images, making it ideal for dynamic, label-free, and lens-free imaging in biological sciences. The GSH is a flexible framework that can be combined with existing regularization methods to improve the noise-robustness of practical phase retrieval tasks, which has broader applications in other phase retrieval tasks such as coherent diffractive imaging \cite{robinson2009coherent,miao2011coherent}, Fourier ptychographic microscopy \cite{zheng2013wide,jiang2023spatial} and diffractive tomography \cite{horstmeyer2016diffraction,zhou2022transport,wu2024lens}, where high-dimensional wavefronts are reconstructed from limited data. The compression strategy in the GSH could be generalized to imaging modalities beyond holography, where the inverse problem is ill-posed, providing a solution to tackle these challenges with reduced dependencies on regularization. Codes that are related to this research are published on Github \cite{GS_github}.


\bibliography{apssamp}

\begin{thebibliography}{49}%
\makeatletter
\providecommand \@ifxundefined [1]{%
 \@ifx{#1\undefined}
}%
\providecommand \@ifnum [1]{%
 \ifnum #1\expandafter \@firstoftwo
 \else \expandafter \@secondoftwo
 \fi
}%
\providecommand \@ifx [1]{%
 \ifx #1\expandafter \@firstoftwo
 \else \expandafter \@secondoftwo
 \fi
}%
\providecommand \natexlab [1]{#1}%
\providecommand \enquote  [1]{``#1''}%
\providecommand \bibnamefont  [1]{#1}%
\providecommand \bibfnamefont [1]{#1}%
\providecommand \citenamefont [1]{#1}%
\providecommand \href@noop [0]{\@secondoftwo}%
\providecommand \href [0]{\begingroup \@sanitize@url \@href}%
\providecommand \@href[1]{\@@startlink{#1}\@@href}%
\providecommand \@@href[1]{\endgroup#1\@@endlink}%
\providecommand \@sanitize@url [0]{\catcode `\\12\catcode `\$12\catcode `\&12\catcode `\#12\catcode `\^12\catcode `\_12\catcode `\%12\relax}%
\providecommand \@@startlink[1]{}%
\providecommand \@@endlink[0]{}%
\providecommand \url  [0]{\begingroup\@sanitize@url \@url }%
\providecommand \@url [1]{\endgroup\@href {#1}{\urlprefix }}%
\providecommand \urlprefix  [0]{URL }%
\providecommand \Eprint [0]{\href }%
\providecommand \doibase [0]{https://doi.org/}%
\providecommand \selectlanguage [0]{\@gobble}%
\providecommand \bibinfo  [0]{\@secondoftwo}%
\providecommand \bibfield  [0]{\@secondoftwo}%
\providecommand \translation [1]{[#1]}%
\providecommand \BibitemOpen [0]{}%
\providecommand \bibitemStop [0]{}%
\providecommand \bibitemNoStop [0]{.\EOS\space}%
\providecommand \EOS [0]{\spacefactor3000\relax}%
\providecommand \BibitemShut  [1]{\csname bibitem#1\endcsname}%
\let\auto@bib@innerbib\@empty
\bibitem [{\citenamefont {Rosen}\ \emph {et~al.}(2024)\citenamefont {Rosen}, \citenamefont {Alford}, \citenamefont {Allan}, \citenamefont {Anand}, \citenamefont {Arnon}, \citenamefont {Arockiaraj}, \citenamefont {Art}, \citenamefont {Bai}, \citenamefont {Balasubramaniam}, \citenamefont {Birnbaum} \emph {et~al.}}]{rosen2024roadmap}%
  \BibitemOpen
  \bibfield  {author} {\bibinfo {author} {\bibfnamefont {J.}~\bibnamefont {Rosen}}, \bibinfo {author} {\bibfnamefont {S.}~\bibnamefont {Alford}}, \bibinfo {author} {\bibfnamefont {B.}~\bibnamefont {Allan}}, \bibinfo {author} {\bibfnamefont {V.}~\bibnamefont {Anand}}, \bibinfo {author} {\bibfnamefont {S.}~\bibnamefont {Arnon}}, \bibinfo {author} {\bibfnamefont {F.~G.}\ \bibnamefont {Arockiaraj}}, \bibinfo {author} {\bibfnamefont {J.}~\bibnamefont {Art}}, \bibinfo {author} {\bibfnamefont {B.}~\bibnamefont {Bai}}, \bibinfo {author} {\bibfnamefont {G.~M.}\ \bibnamefont {Balasubramaniam}}, \bibinfo {author} {\bibfnamefont {T.}~\bibnamefont {Birnbaum}}, \emph {et~al.},\ }\bibfield  {title} {\bibinfo {title} {Roadmap on computational methods in optical imaging and holography},\ }\href@noop {} {\bibfield  {journal} {\bibinfo  {journal} {Applied Physics B}\ }\textbf {\bibinfo {volume} {130}},\ \bibinfo {pages} {166} (\bibinfo {year} {2024})}\BibitemShut {NoStop}%
\bibitem [{\citenamefont {Mann}\ \emph {et~al.}(2005)\citenamefont {Mann}, \citenamefont {Yu}, \citenamefont {Lo},\ and\ \citenamefont {Kim}}]{mann2005high}%
  \BibitemOpen
  \bibfield  {author} {\bibinfo {author} {\bibfnamefont {C.~J.}\ \bibnamefont {Mann}}, \bibinfo {author} {\bibfnamefont {L.}~\bibnamefont {Yu}}, \bibinfo {author} {\bibfnamefont {C.-M.}\ \bibnamefont {Lo}},\ and\ \bibinfo {author} {\bibfnamefont {M.~K.}\ \bibnamefont {Kim}},\ }\bibfield  {title} {\bibinfo {title} {High-resolution quantitative phase-contrast microscopy by digital holography},\ }\href@noop {} {\bibfield  {journal} {\bibinfo  {journal} {Optics Express}\ }\textbf {\bibinfo {volume} {13}},\ \bibinfo {pages} {8693} (\bibinfo {year} {2005})}\BibitemShut {NoStop}%
\bibitem [{\citenamefont {Huang}\ and\ \citenamefont {Cao}(2024)}]{huang2024quantitative}%
  \BibitemOpen
  \bibfield  {author} {\bibinfo {author} {\bibfnamefont {Z.}~\bibnamefont {Huang}}\ and\ \bibinfo {author} {\bibfnamefont {L.}~\bibnamefont {Cao}},\ }\bibfield  {title} {\bibinfo {title} {Quantitative phase imaging based on holography: trends and new perspectives},\ }\href@noop {} {\bibfield  {journal} {\bibinfo  {journal} {Light: Science \& Applications}\ }\textbf {\bibinfo {volume} {13}},\ \bibinfo {pages} {145} (\bibinfo {year} {2024})}\BibitemShut {NoStop}%
\bibitem [{\citenamefont {Kim}\ \emph {et~al.}(2014)\citenamefont {Kim}, \citenamefont {Zhou}, \citenamefont {Mir}, \citenamefont {Babacan}, \citenamefont {Carney}, \citenamefont {Goddard},\ and\ \citenamefont {Popescu}}]{kim2014white}%
  \BibitemOpen
  \bibfield  {author} {\bibinfo {author} {\bibfnamefont {T.}~\bibnamefont {Kim}}, \bibinfo {author} {\bibfnamefont {R.}~\bibnamefont {Zhou}}, \bibinfo {author} {\bibfnamefont {M.}~\bibnamefont {Mir}}, \bibinfo {author} {\bibfnamefont {S.~D.}\ \bibnamefont {Babacan}}, \bibinfo {author} {\bibfnamefont {P.~S.}\ \bibnamefont {Carney}}, \bibinfo {author} {\bibfnamefont {L.~L.}\ \bibnamefont {Goddard}},\ and\ \bibinfo {author} {\bibfnamefont {G.}~\bibnamefont {Popescu}},\ }\bibfield  {title} {\bibinfo {title} {White-light diffraction tomography of unlabelled live cells},\ }\href@noop {} {\bibfield  {journal} {\bibinfo  {journal} {Nature Photonics}\ }\textbf {\bibinfo {volume} {8}},\ \bibinfo {pages} {256} (\bibinfo {year} {2014})}\BibitemShut {NoStop}%
\bibitem [{\citenamefont {Shi}\ \emph {et~al.}(2021)\citenamefont {Shi}, \citenamefont {Li}, \citenamefont {Kim}, \citenamefont {Kellnhofer},\ and\ \citenamefont {Matusik}}]{shi2021towards}%
  \BibitemOpen
  \bibfield  {author} {\bibinfo {author} {\bibfnamefont {L.}~\bibnamefont {Shi}}, \bibinfo {author} {\bibfnamefont {B.}~\bibnamefont {Li}}, \bibinfo {author} {\bibfnamefont {C.}~\bibnamefont {Kim}}, \bibinfo {author} {\bibfnamefont {P.}~\bibnamefont {Kellnhofer}},\ and\ \bibinfo {author} {\bibfnamefont {W.}~\bibnamefont {Matusik}},\ }\bibfield  {title} {\bibinfo {title} {Towards real-time photorealistic {3D} holography with deep neural networks},\ }\href@noop {} {\bibfield  {journal} {\bibinfo  {journal} {Nature}\ }\textbf {\bibinfo {volume} {591}},\ \bibinfo {pages} {234} (\bibinfo {year} {2021})}\BibitemShut {NoStop}%
\bibitem [{\citenamefont {Paturzo}\ \emph {et~al.}(2018)\citenamefont {Paturzo}, \citenamefont {Pagliarulo}, \citenamefont {Bianco}, \citenamefont {Memmolo}, \citenamefont {Miccio}, \citenamefont {Merola},\ and\ \citenamefont {Ferraro}}]{paturzo2018digital}%
  \BibitemOpen
  \bibfield  {author} {\bibinfo {author} {\bibfnamefont {M.}~\bibnamefont {Paturzo}}, \bibinfo {author} {\bibfnamefont {V.}~\bibnamefont {Pagliarulo}}, \bibinfo {author} {\bibfnamefont {V.}~\bibnamefont {Bianco}}, \bibinfo {author} {\bibfnamefont {P.}~\bibnamefont {Memmolo}}, \bibinfo {author} {\bibfnamefont {L.}~\bibnamefont {Miccio}}, \bibinfo {author} {\bibfnamefont {F.}~\bibnamefont {Merola}},\ and\ \bibinfo {author} {\bibfnamefont {P.}~\bibnamefont {Ferraro}},\ }\bibfield  {title} {\bibinfo {title} {Digital holography, a metrological tool for quantitative analysis: Trends and future applications},\ }\href@noop {} {\bibfield  {journal} {\bibinfo  {journal} {Optics and Lasers in Engineering}\ }\textbf {\bibinfo {volume} {104}},\ \bibinfo {pages} {32} (\bibinfo {year} {2018})}\BibitemShut {NoStop}%
\bibitem [{\citenamefont {Gabor}(1948)}]{Gabor1948}%
  \BibitemOpen
  \bibfield  {author} {\bibinfo {author} {\bibfnamefont {D.}~\bibnamefont {Gabor}},\ }\bibfield  {title} {\bibinfo {title} {A new microscopic principle},\ }\href@noop {} {\bibfield  {journal} {\bibinfo  {journal} {Nature}\ }\textbf {\bibinfo {volume} {161}},\ \bibinfo {pages} {777} (\bibinfo {year} {1948})}\BibitemShut {NoStop}%
\bibitem [{\citenamefont {Latychevskaia}\ and\ \citenamefont {Fink}(2007)}]{PhysRevLett.98.233901}%
  \BibitemOpen
  \bibfield  {author} {\bibinfo {author} {\bibfnamefont {T.}~\bibnamefont {Latychevskaia}}\ and\ \bibinfo {author} {\bibfnamefont {H.-W.}\ \bibnamefont {Fink}},\ }\bibfield  {title} {\bibinfo {title} {Solution to the twin image problem in holography},\ }\href@noop {} {\bibfield  {journal} {\bibinfo  {journal} {Phys. Rev. Lett.}\ }\textbf {\bibinfo {volume} {98}},\ \bibinfo {pages} {233901} (\bibinfo {year} {2007})}\BibitemShut {NoStop}%
\bibitem [{\citenamefont {Zhang}\ \emph {et~al.}(2018{\natexlab{a}})\citenamefont {Zhang}, \citenamefont {Cao}, \citenamefont {Brady}, \citenamefont {Zhang}, \citenamefont {Cang}, \citenamefont {Zhang},\ and\ \citenamefont {Jin}}]{PhysRevLett.121.093902}%
  \BibitemOpen
  \bibfield  {author} {\bibinfo {author} {\bibfnamefont {W.}~\bibnamefont {Zhang}}, \bibinfo {author} {\bibfnamefont {L.}~\bibnamefont {Cao}}, \bibinfo {author} {\bibfnamefont {D.~J.}\ \bibnamefont {Brady}}, \bibinfo {author} {\bibfnamefont {H.}~\bibnamefont {Zhang}}, \bibinfo {author} {\bibfnamefont {J.}~\bibnamefont {Cang}}, \bibinfo {author} {\bibfnamefont {H.}~\bibnamefont {Zhang}},\ and\ \bibinfo {author} {\bibfnamefont {G.}~\bibnamefont {Jin}},\ }\bibfield  {title} {\bibinfo {title} {Twin-image-free holography: A compressive sensing approach},\ }\href@noop {} {\bibfield  {journal} {\bibinfo  {journal} {Phys. Rev. Lett.}\ }\textbf {\bibinfo {volume} {121}},\ \bibinfo {pages} {093902} (\bibinfo {year} {2018}{\natexlab{a}})}\BibitemShut {NoStop}%
\bibitem [{\citenamefont {Wu}\ \emph {et~al.}(2020)\citenamefont {Wu}, \citenamefont {Zhang}, \citenamefont {Zhang}, \citenamefont {Jin}, \citenamefont {Cao},\ and\ \citenamefont {Barbastathis}}]{wu2020single}%
  \BibitemOpen
  \bibfield  {author} {\bibinfo {author} {\bibfnamefont {J.}~\bibnamefont {Wu}}, \bibinfo {author} {\bibfnamefont {H.}~\bibnamefont {Zhang}}, \bibinfo {author} {\bibfnamefont {W.}~\bibnamefont {Zhang}}, \bibinfo {author} {\bibfnamefont {G.}~\bibnamefont {Jin}}, \bibinfo {author} {\bibfnamefont {L.}~\bibnamefont {Cao}},\ and\ \bibinfo {author} {\bibfnamefont {G.}~\bibnamefont {Barbastathis}},\ }\bibfield  {title} {\bibinfo {title} {Single-shot lensless imaging with fresnel zone aperture and incoherent illumination},\ }\href@noop {} {\bibfield  {journal} {\bibinfo  {journal} {Light: Science \& Applications}\ }\textbf {\bibinfo {volume} {9}},\ \bibinfo {pages} {53} (\bibinfo {year} {2020})}\BibitemShut {NoStop}%
\bibitem [{\citenamefont {Gao}\ and\ \citenamefont {Cao}(2023)}]{gao2023iterative}%
  \BibitemOpen
  \bibfield  {author} {\bibinfo {author} {\bibfnamefont {Y.}~\bibnamefont {Gao}}\ and\ \bibinfo {author} {\bibfnamefont {L.}~\bibnamefont {Cao}},\ }\bibfield  {title} {\bibinfo {title} {Iterative projection meets sparsity regularization: towards practical single-shot quantitative phase imaging with in-line holography},\ }\href@noop {} {\bibfield  {journal} {\bibinfo  {journal} {Light: Advanced Manufacturing}\ }\textbf {\bibinfo {volume} {4}},\ \bibinfo {pages} {37} (\bibinfo {year} {2023})}\BibitemShut {NoStop}%
\bibitem [{\citenamefont {Popescu}(2021)}]{popescu2021large}%
  \BibitemOpen
  \bibfield  {author} {\bibinfo {author} {\bibfnamefont {G.}~\bibnamefont {Popescu}},\ }\bibfield  {title} {\bibinfo {title} {Large-scale phase retrieval},\ }\href@noop {} {\bibfield  {journal} {\bibinfo  {journal} {Light, Science \& Applications}\ }\textbf {\bibinfo {volume} {10}} (\bibinfo {year} {2021})}\BibitemShut {NoStop}%
\bibitem [{\citenamefont {Paxman}\ \emph {et~al.}(1992)\citenamefont {Paxman}, \citenamefont {Schulz},\ and\ \citenamefont {Fienup}}]{paxman1992joint}%
  \BibitemOpen
  \bibfield  {author} {\bibinfo {author} {\bibfnamefont {R.~G.}\ \bibnamefont {Paxman}}, \bibinfo {author} {\bibfnamefont {T.~J.}\ \bibnamefont {Schulz}},\ and\ \bibinfo {author} {\bibfnamefont {J.~R.}\ \bibnamefont {Fienup}},\ }\bibfield  {title} {\bibinfo {title} {Joint estimation of object and aberrations by using phase diversity},\ }\href@noop {} {\bibfield  {journal} {\bibinfo  {journal} {Journal of the Optical Society of America A}\ }\textbf {\bibinfo {volume} {9}},\ \bibinfo {pages} {1072} (\bibinfo {year} {1992})}\BibitemShut {NoStop}%
\bibitem [{\citenamefont {Rodenburg}\ and\ \citenamefont {Faulkner}(2004)}]{rodenburg2004phase}%
  \BibitemOpen
  \bibfield  {author} {\bibinfo {author} {\bibfnamefont {J.~M.}\ \bibnamefont {Rodenburg}}\ and\ \bibinfo {author} {\bibfnamefont {H.~M.}\ \bibnamefont {Faulkner}},\ }\bibfield  {title} {\bibinfo {title} {A phase retrieval algorithm for shifting illumination},\ }\href@noop {} {\bibfield  {journal} {\bibinfo  {journal} {Applied Physics Letters}\ }\textbf {\bibinfo {volume} {85}},\ \bibinfo {pages} {4795} (\bibinfo {year} {2004})}\BibitemShut {NoStop}%
\bibitem [{\citenamefont {Jiang}\ \emph {et~al.}(2023)\citenamefont {Jiang}, \citenamefont {Song}, \citenamefont {Wang}, \citenamefont {Yang}, \citenamefont {Wang}, \citenamefont {Guo}, \citenamefont {Feng}, \citenamefont {Maiden},\ and\ \citenamefont {Zheng}}]{jiang2023spatial}%
  \BibitemOpen
  \bibfield  {author} {\bibinfo {author} {\bibfnamefont {S.}~\bibnamefont {Jiang}}, \bibinfo {author} {\bibfnamefont {P.}~\bibnamefont {Song}}, \bibinfo {author} {\bibfnamefont {T.}~\bibnamefont {Wang}}, \bibinfo {author} {\bibfnamefont {L.}~\bibnamefont {Yang}}, \bibinfo {author} {\bibfnamefont {R.}~\bibnamefont {Wang}}, \bibinfo {author} {\bibfnamefont {C.}~\bibnamefont {Guo}}, \bibinfo {author} {\bibfnamefont {B.}~\bibnamefont {Feng}}, \bibinfo {author} {\bibfnamefont {A.}~\bibnamefont {Maiden}},\ and\ \bibinfo {author} {\bibfnamefont {G.}~\bibnamefont {Zheng}},\ }\bibfield  {title} {\bibinfo {title} {Spatial-and {F}ourier-domain ptychography for high-throughput bio-imaging},\ }\href@noop {} {\bibfield  {journal} {\bibinfo  {journal} {Nature Protocols}\ }\textbf {\bibinfo {volume} {18}},\ \bibinfo {pages} {2051} (\bibinfo {year} {2023})}\BibitemShut {NoStop}%
\bibitem [{\citenamefont {Brady}\ \emph {et~al.}(2009)\citenamefont {Brady}, \citenamefont {Choi}, \citenamefont {Marks}, \citenamefont {Horisaki},\ and\ \citenamefont {Lim}}]{Brady:09}%
  \BibitemOpen
  \bibfield  {author} {\bibinfo {author} {\bibfnamefont {D.~J.}\ \bibnamefont {Brady}}, \bibinfo {author} {\bibfnamefont {K.}~\bibnamefont {Choi}}, \bibinfo {author} {\bibfnamefont {D.~L.}\ \bibnamefont {Marks}}, \bibinfo {author} {\bibfnamefont {R.}~\bibnamefont {Horisaki}},\ and\ \bibinfo {author} {\bibfnamefont {S.}~\bibnamefont {Lim}},\ }\bibfield  {title} {\bibinfo {title} {Compressive holography},\ }\href@noop {} {\bibfield  {journal} {\bibinfo  {journal} {Optics Express}\ }\textbf {\bibinfo {volume} {17}},\ \bibinfo {pages} {13040} (\bibinfo {year} {2009})}\BibitemShut {NoStop}%
\bibitem [{\citenamefont {Rivenson}\ \emph {et~al.}(2019)\citenamefont {Rivenson}, \citenamefont {Wu},\ and\ \citenamefont {Ozcan}}]{rivenson2019deep}%
  \BibitemOpen
  \bibfield  {author} {\bibinfo {author} {\bibfnamefont {Y.}~\bibnamefont {Rivenson}}, \bibinfo {author} {\bibfnamefont {Y.}~\bibnamefont {Wu}},\ and\ \bibinfo {author} {\bibfnamefont {A.}~\bibnamefont {Ozcan}},\ }\bibfield  {title} {\bibinfo {title} {Deep learning in holography and coherent imaging},\ }\href@noop {} {\bibfield  {journal} {\bibinfo  {journal} {Light: Science \& Applications}\ }\textbf {\bibinfo {volume} {8}},\ \bibinfo {pages} {85} (\bibinfo {year} {2019})}\BibitemShut {NoStop}%
\bibitem [{\citenamefont {Wang}\ \emph {et~al.}(2020)\citenamefont {Wang}, \citenamefont {Bian}, \citenamefont {Wang}, \citenamefont {Lyu}, \citenamefont {Pedrini}, \citenamefont {Osten}, \citenamefont {Barbastathis},\ and\ \citenamefont {Situ}}]{wang2020phase}%
  \BibitemOpen
  \bibfield  {author} {\bibinfo {author} {\bibfnamefont {F.}~\bibnamefont {Wang}}, \bibinfo {author} {\bibfnamefont {Y.}~\bibnamefont {Bian}}, \bibinfo {author} {\bibfnamefont {H.}~\bibnamefont {Wang}}, \bibinfo {author} {\bibfnamefont {M.}~\bibnamefont {Lyu}}, \bibinfo {author} {\bibfnamefont {G.}~\bibnamefont {Pedrini}}, \bibinfo {author} {\bibfnamefont {W.}~\bibnamefont {Osten}}, \bibinfo {author} {\bibfnamefont {G.}~\bibnamefont {Barbastathis}},\ and\ \bibinfo {author} {\bibfnamefont {G.}~\bibnamefont {Situ}},\ }\bibfield  {title} {\bibinfo {title} {Phase imaging with an untrained neural network},\ }\href@noop {} {\bibfield  {journal} {\bibinfo  {journal} {Light: Science \& Applications}\ }\textbf {\bibinfo {volume} {9}},\ \bibinfo {pages} {77} (\bibinfo {year} {2020})}\BibitemShut {NoStop}%
\bibitem [{\citenamefont {Chang}\ \emph {et~al.}(2021)\citenamefont {Chang}, \citenamefont {Bian},\ and\ \citenamefont {Zhang}}]{chang2021large}%
  \BibitemOpen
  \bibfield  {author} {\bibinfo {author} {\bibfnamefont {X.}~\bibnamefont {Chang}}, \bibinfo {author} {\bibfnamefont {L.}~\bibnamefont {Bian}},\ and\ \bibinfo {author} {\bibfnamefont {J.}~\bibnamefont {Zhang}},\ }\bibfield  {title} {\bibinfo {title} {Large-scale phase retrieval},\ }\href@noop {} {\bibfield  {journal} {\bibinfo  {journal} {Elight}\ }\textbf {\bibinfo {volume} {1}},\ \bibinfo {pages} {4} (\bibinfo {year} {2021})}\BibitemShut {NoStop}%
\bibitem [{\citenamefont {Kerbl}\ \emph {et~al.}(2023)\citenamefont {Kerbl}, \citenamefont {Kopanas}, \citenamefont {Leimk{\"u}hler},\ and\ \citenamefont {Drettakis}}]{KKLD23}%
  \BibitemOpen
  \bibfield  {author} {\bibinfo {author} {\bibfnamefont {B.}~\bibnamefont {Kerbl}}, \bibinfo {author} {\bibfnamefont {G.}~\bibnamefont {Kopanas}}, \bibinfo {author} {\bibfnamefont {T.}~\bibnamefont {Leimk{\"u}hler}},\ and\ \bibinfo {author} {\bibfnamefont {G.}~\bibnamefont {Drettakis}},\ }\bibfield  {title} {\bibinfo {title} {{3D} {G}aussian splatting for real-time radiance field rendering},\ }\href@noop {} {\bibfield  {journal} {\bibinfo  {journal} {ACM Transactions on Graphics (SIGGRAPH Conference Proceedings)}\ }\textbf {\bibinfo {volume} {42}} (\bibinfo {year} {2023})}\BibitemShut {NoStop}%
\bibitem [{\citenamefont {Zhang}\ \emph {et~al.}(2025)\citenamefont {Zhang}, \citenamefont {Ge}, \citenamefont {Xu}, \citenamefont {He}, \citenamefont {Wang}, \citenamefont {Qin}, \citenamefont {Lu}, \citenamefont {Geng},\ and\ \citenamefont {Zhang}}]{GaussianImage}%
  \BibitemOpen
  \bibfield  {author} {\bibinfo {author} {\bibfnamefont {X.}~\bibnamefont {Zhang}}, \bibinfo {author} {\bibfnamefont {X.}~\bibnamefont {Ge}}, \bibinfo {author} {\bibfnamefont {T.}~\bibnamefont {Xu}}, \bibinfo {author} {\bibfnamefont {D.}~\bibnamefont {He}}, \bibinfo {author} {\bibfnamefont {Y.}~\bibnamefont {Wang}}, \bibinfo {author} {\bibfnamefont {H.}~\bibnamefont {Qin}}, \bibinfo {author} {\bibfnamefont {G.}~\bibnamefont {Lu}}, \bibinfo {author} {\bibfnamefont {J.}~\bibnamefont {Geng}},\ and\ \bibinfo {author} {\bibfnamefont {J.}~\bibnamefont {Zhang}},\ }in\ \href@noop {} {\emph {\bibinfo {booktitle} {Computer Vision -- ECCV 2024}}}\ (\bibinfo  {publisher} {Springer Nature Switzerland},\ \bibinfo {address} {Cham},\ \bibinfo {year} {2025})\ pp.\ \bibinfo {pages} {327--345}\BibitemShut {NoStop}%
\bibitem [{\citenamefont {Niedermayr}\ \emph {et~al.}(2024)\citenamefont {Niedermayr}, \citenamefont {Stumpfegger},\ and\ \citenamefont {Westermann}}]{niedermayr2024compressed}%
  \BibitemOpen
  \bibfield  {author} {\bibinfo {author} {\bibfnamefont {S.}~\bibnamefont {Niedermayr}}, \bibinfo {author} {\bibfnamefont {J.}~\bibnamefont {Stumpfegger}},\ and\ \bibinfo {author} {\bibfnamefont {R.}~\bibnamefont {Westermann}},\ }\bibfield  {title} {\bibinfo {title} {Compressed {3D Gaussian} splatting for accelerated novel view synthesis},\ }in\ \href@noop {} {\emph {\bibinfo {booktitle} {Proceedings of the IEEE/CVF Conference on Computer Vision and Pattern Recognition}}}\ (\bibinfo {year} {2024})\ pp.\ \bibinfo {pages} {10349--10358}\BibitemShut {NoStop}%
\bibitem [{\citenamefont {Chen}\ \emph {et~al.}(2024{\natexlab{a}})\citenamefont {Chen}, \citenamefont {Wu}, \citenamefont {Lin}, \citenamefont {Harandi},\ and\ \citenamefont {Cai}}]{chen2024hac}%
  \BibitemOpen
  \bibfield  {author} {\bibinfo {author} {\bibfnamefont {Y.}~\bibnamefont {Chen}}, \bibinfo {author} {\bibfnamefont {Q.}~\bibnamefont {Wu}}, \bibinfo {author} {\bibfnamefont {W.}~\bibnamefont {Lin}}, \bibinfo {author} {\bibfnamefont {M.}~\bibnamefont {Harandi}},\ and\ \bibinfo {author} {\bibfnamefont {J.}~\bibnamefont {Cai}},\ }\bibfield  {title} {\bibinfo {title} {Hac: Hash-grid assisted context for {3D} {G}aussian splatting compression},\ }in\ \href@noop {} {\emph {\bibinfo {booktitle} {European Conference on Computer Vision}}}\ (\bibinfo {organization} {Springer},\ \bibinfo {year} {2024})\ pp.\ \bibinfo {pages} {422--438}\BibitemShut {NoStop}%
\bibitem [{\citenamefont {Carter}(1972)}]{carter1972electromagnetic}%
  \BibitemOpen
  \bibfield  {author} {\bibinfo {author} {\bibfnamefont {W.~H.}\ \bibnamefont {Carter}},\ }\bibfield  {title} {\bibinfo {title} {Electromagnetic field of a {G}aussian beam with an elliptical cross section},\ }\href@noop {} {\bibfield  {journal} {\bibinfo  {journal} {Journal of the Optical Society of America}\ }\textbf {\bibinfo {volume} {62}},\ \bibinfo {pages} {1195} (\bibinfo {year} {1972})}\BibitemShut {NoStop}%
\bibitem [{\citenamefont {Cornolti}\ \emph {et~al.}(1990)\citenamefont {Cornolti}, \citenamefont {Lucchesi},\ and\ \citenamefont {Zambon}}]{cornolti1990elliptic}%
  \BibitemOpen
  \bibfield  {author} {\bibinfo {author} {\bibfnamefont {F.}~\bibnamefont {Cornolti}}, \bibinfo {author} {\bibfnamefont {M.}~\bibnamefont {Lucchesi}},\ and\ \bibinfo {author} {\bibfnamefont {B.}~\bibnamefont {Zambon}},\ }\bibfield  {title} {\bibinfo {title} {Elliptic {G}aussian beam self-focusing in nonlinear media},\ }\href@noop {} {\bibfield  {journal} {\bibinfo  {journal} {Optics Communications}\ }\textbf {\bibinfo {volume} {75}},\ \bibinfo {pages} {129} (\bibinfo {year} {1990})}\BibitemShut {NoStop}%
\bibitem [{\citenamefont {Heyman}\ and\ \citenamefont {Felsen}(2001)}]{heyman2001gaussian}%
  \BibitemOpen
  \bibfield  {author} {\bibinfo {author} {\bibfnamefont {E.}~\bibnamefont {Heyman}}\ and\ \bibinfo {author} {\bibfnamefont {L.~B.}\ \bibnamefont {Felsen}},\ }\bibfield  {title} {\bibinfo {title} {{G}aussian beam and pulsed-beam dynamics: complex-source and complex-spectrum formulations within and beyond paraxial asymptotics},\ }\href@noop {} {\bibfield  {journal} {\bibinfo  {journal} {Journal of the Optical Society of America A}\ }\textbf {\bibinfo {volume} {18}},\ \bibinfo {pages} {1588} (\bibinfo {year} {2001})}\BibitemShut {NoStop}%
\bibitem [{\citenamefont {Cai}\ and\ \citenamefont {Lin}(2002)}]{cai2002decentered}%
  \BibitemOpen
  \bibfield  {author} {\bibinfo {author} {\bibfnamefont {Y.}~\bibnamefont {Cai}}\ and\ \bibinfo {author} {\bibfnamefont {Q.}~\bibnamefont {Lin}},\ }\bibfield  {title} {\bibinfo {title} {Decentered elliptical {G}aussian beam},\ }\href@noop {} {\bibfield  {journal} {\bibinfo  {journal} {Applied Optics}\ }\textbf {\bibinfo {volume} {41}},\ \bibinfo {pages} {4336} (\bibinfo {year} {2002})}\BibitemShut {NoStop}%
\bibitem [{\citenamefont {Dickson}(1970)}]{dickson1970characteristics}%
  \BibitemOpen
  \bibfield  {author} {\bibinfo {author} {\bibfnamefont {L.~D.}\ \bibnamefont {Dickson}},\ }\bibfield  {title} {\bibinfo {title} {Characteristics of a propagating {G}aussian beam},\ }\href@noop {} {\bibfield  {journal} {\bibinfo  {journal} {Applied Optics}\ }\textbf {\bibinfo {volume} {9}},\ \bibinfo {pages} {1854} (\bibinfo {year} {1970})}\BibitemShut {NoStop}%
\bibitem [{\citenamefont {Agrawal}\ and\ \citenamefont {Pattanayak}(1979)}]{agrawal1979gaussian}%
  \BibitemOpen
  \bibfield  {author} {\bibinfo {author} {\bibfnamefont {G.~P.}\ \bibnamefont {Agrawal}}\ and\ \bibinfo {author} {\bibfnamefont {D.~N.}\ \bibnamefont {Pattanayak}},\ }\bibfield  {title} {\bibinfo {title} {{G}aussian beam propagation beyond the paraxial approximation},\ }\href@noop {} {\bibfield  {journal} {\bibinfo  {journal} {Journal of the Optical Society of America}\ }\textbf {\bibinfo {volume} {69}},\ \bibinfo {pages} {575} (\bibinfo {year} {1979})}\BibitemShut {NoStop}%
\bibitem [{\citenamefont {Alda}(2003)}]{alda2003laser}%
  \BibitemOpen
  \bibfield  {author} {\bibinfo {author} {\bibfnamefont {J.}~\bibnamefont {Alda}},\ }\bibfield  {title} {\bibinfo {title} {Laser and {G}aussian beam propagation and transformation},\ }\href@noop {} {\bibfield  {journal} {\bibinfo  {journal} {Encyclopedia of Optical Engineering}\ }\textbf {\bibinfo {volume} {999}},\ \bibinfo {pages} {1013} (\bibinfo {year} {2003})}\BibitemShut {NoStop}%
\bibitem [{\citenamefont {Zhang}\ \emph {et~al.}(2018{\natexlab{b}})\citenamefont {Zhang}, \citenamefont {Zhou},\ and\ \citenamefont {Gong}}]{zhang2018skew}%
  \BibitemOpen
  \bibfield  {author} {\bibinfo {author} {\bibfnamefont {S.}~\bibnamefont {Zhang}}, \bibinfo {author} {\bibfnamefont {J.}~\bibnamefont {Zhou}},\ and\ \bibinfo {author} {\bibfnamefont {L.}~\bibnamefont {Gong}},\ }\bibfield  {title} {\bibinfo {title} {Skew line ray model of nonparaxial {G}aussian beam},\ }\href@noop {} {\bibfield  {journal} {\bibinfo  {journal} {Optics Express}\ }\textbf {\bibinfo {volume} {26}},\ \bibinfo {pages} {3381} (\bibinfo {year} {2018}{\natexlab{b}})}\BibitemShut {NoStop}%
\bibitem [{\citenamefont {Goodman}(2005)}]{goodman2005introduction}%
  \BibitemOpen
  \bibfield  {author} {\bibinfo {author} {\bibfnamefont {J.~W.}\ \bibnamefont {Goodman}},\ }\href@noop {} {\emph {\bibinfo {title} {Introduction to {F}ourier optics}}}\ (\bibinfo  {publisher} {Roberts and Company publishers},\ \bibinfo {year} {2005})\BibitemShut {NoStop}%
\bibitem [{\citenamefont {Chen}\ \emph {et~al.}(2024{\natexlab{b}})\citenamefont {Chen}, \citenamefont {Liang}, \citenamefont {Huang}, \citenamefont {Real}, \citenamefont {Wang}, \citenamefont {Pham}, \citenamefont {Dong}, \citenamefont {Luong}, \citenamefont {Hsieh}, \citenamefont {Lu} \emph {et~al.}}]{chen2024symbolic}%
  \BibitemOpen
  \bibfield  {author} {\bibinfo {author} {\bibfnamefont {X.}~\bibnamefont {Chen}}, \bibinfo {author} {\bibfnamefont {C.}~\bibnamefont {Liang}}, \bibinfo {author} {\bibfnamefont {D.}~\bibnamefont {Huang}}, \bibinfo {author} {\bibfnamefont {E.}~\bibnamefont {Real}}, \bibinfo {author} {\bibfnamefont {K.}~\bibnamefont {Wang}}, \bibinfo {author} {\bibfnamefont {H.}~\bibnamefont {Pham}}, \bibinfo {author} {\bibfnamefont {X.}~\bibnamefont {Dong}}, \bibinfo {author} {\bibfnamefont {T.}~\bibnamefont {Luong}}, \bibinfo {author} {\bibfnamefont {C.-J.}\ \bibnamefont {Hsieh}}, \bibinfo {author} {\bibfnamefont {Y.}~\bibnamefont {Lu}}, \emph {et~al.},\ }\bibfield  {title} {\bibinfo {title} {Symbolic discovery of optimization algorithms},\ }\href@noop {} {\bibfield  {journal} {\bibinfo  {journal} {Advances in Neural Information Processing Systems}\ }\textbf {\bibinfo {volume} {36}} (\bibinfo {year} {2024}{\natexlab{b}})}\BibitemShut {NoStop}%
\bibitem [{\citenamefont {Ye}\ \emph {et~al.}(2024)\citenamefont {Ye}, \citenamefont {Li}, \citenamefont {Kerr}, \citenamefont {Turkulainen}, \citenamefont {Yi}, \citenamefont {Pan}, \citenamefont {Seiskari}, \citenamefont {Ye}, \citenamefont {Hu}, \citenamefont {Tancik} \emph {et~al.}}]{ye2024gsplat}%
  \BibitemOpen
  \bibfield  {author} {\bibinfo {author} {\bibfnamefont {V.}~\bibnamefont {Ye}}, \bibinfo {author} {\bibfnamefont {R.}~\bibnamefont {Li}}, \bibinfo {author} {\bibfnamefont {J.}~\bibnamefont {Kerr}}, \bibinfo {author} {\bibfnamefont {M.}~\bibnamefont {Turkulainen}}, \bibinfo {author} {\bibfnamefont {B.}~\bibnamefont {Yi}}, \bibinfo {author} {\bibfnamefont {Z.}~\bibnamefont {Pan}}, \bibinfo {author} {\bibfnamefont {O.}~\bibnamefont {Seiskari}}, \bibinfo {author} {\bibfnamefont {J.}~\bibnamefont {Ye}}, \bibinfo {author} {\bibfnamefont {J.}~\bibnamefont {Hu}}, \bibinfo {author} {\bibfnamefont {M.}~\bibnamefont {Tancik}}, \emph {et~al.},\ }\bibfield  {title} {\bibinfo {title} {gsplat: An open-source library for {G}aussian splatting},\ }\href@noop {} {\bibfield  {journal} {\bibinfo  {journal} {arXiv preprint arXiv:2409.06765}\ } (\bibinfo {year} {2024})}\BibitemShut {NoStop}%
\bibitem [{\citenamefont {Elser}(2003)}]{Elser:03}%
  \BibitemOpen
  \bibfield  {author} {\bibinfo {author} {\bibfnamefont {V.}~\bibnamefont {Elser}},\ }\bibfield  {title} {\bibinfo {title} {Phase retrieval by iterated projections},\ }\href@noop {} {\bibfield  {journal} {\bibinfo  {journal} {Journal of the Optical Society of America A}\ }\textbf {\bibinfo {volume} {20}},\ \bibinfo {pages} {40} (\bibinfo {year} {2003})}\BibitemShut {NoStop}%
\bibitem [{\citenamefont {Jaganathan}\ \emph {et~al.}(2016)\citenamefont {Jaganathan}, \citenamefont {Eldar},\ and\ \citenamefont {Hassibi}}]{jaganathan2016phase}%
  \BibitemOpen
  \bibfield  {author} {\bibinfo {author} {\bibfnamefont {K.}~\bibnamefont {Jaganathan}}, \bibinfo {author} {\bibfnamefont {Y.~C.}\ \bibnamefont {Eldar}},\ and\ \bibinfo {author} {\bibfnamefont {B.}~\bibnamefont {Hassibi}},\ }\bibfield  {title} {\bibinfo {title} {Phase retrieval: An overview of recent developments},\ }\href@noop {} {\bibfield  {journal} {\bibinfo  {journal} {Optical compressive imaging}\ ,\ \bibinfo {pages} {279}} (\bibinfo {year} {2016})}\BibitemShut {NoStop}%
\bibitem [{\citenamefont {Hore}\ and\ \citenamefont {Ziou}(2010)}]{hore2010image}%
  \BibitemOpen
  \bibfield  {author} {\bibinfo {author} {\bibfnamefont {A.}~\bibnamefont {Hore}}\ and\ \bibinfo {author} {\bibfnamefont {D.}~\bibnamefont {Ziou}},\ }\bibfield  {title} {\bibinfo {title} {Image quality metrics: Psnr vs. ssim},\ }in\ \href@noop {} {\emph {\bibinfo {booktitle} {2010 20th international conference on pattern recognition}}}\ (\bibinfo {organization} {IEEE},\ \bibinfo {year} {2010})\ pp.\ \bibinfo {pages} {2366--2369}\BibitemShut {NoStop}%
\bibitem [{\citenamefont {Rudin}\ \emph {et~al.}(1992)\citenamefont {Rudin}, \citenamefont {Osher},\ and\ \citenamefont {Fatemi}}]{rudin1992nonlinear}%
  \BibitemOpen
  \bibfield  {author} {\bibinfo {author} {\bibfnamefont {L.~I.}\ \bibnamefont {Rudin}}, \bibinfo {author} {\bibfnamefont {S.}~\bibnamefont {Osher}},\ and\ \bibinfo {author} {\bibfnamefont {E.}~\bibnamefont {Fatemi}},\ }\bibfield  {title} {\bibinfo {title} {Nonlinear total variation based noise removal algorithms},\ }\href@noop {} {\bibfield  {journal} {\bibinfo  {journal} {Physica D: nonlinear phenomena}\ }\textbf {\bibinfo {volume} {60}},\ \bibinfo {pages} {259} (\bibinfo {year} {1992})}\BibitemShut {NoStop}%
\bibitem [{\citenamefont {Rudin}\ and\ \citenamefont {Osher}(1994)}]{rudin1994total}%
  \BibitemOpen
  \bibfield  {author} {\bibinfo {author} {\bibfnamefont {L.~I.}\ \bibnamefont {Rudin}}\ and\ \bibinfo {author} {\bibfnamefont {S.}~\bibnamefont {Osher}},\ }\bibfield  {title} {\bibinfo {title} {Total variation based image restoration with free local constraints},\ }in\ \href@noop {} {\emph {\bibinfo {booktitle} {Proceedings of 1st international conference on image processing}}},\ Vol.~\bibinfo {volume} {1}\ (\bibinfo {organization} {IEEE},\ \bibinfo {year} {1994})\ pp.\ \bibinfo {pages} {31--35}\BibitemShut {NoStop}%
\bibitem [{\citenamefont {Chambolle}(2004)}]{chambolle2004algorithm}%
  \BibitemOpen
  \bibfield  {author} {\bibinfo {author} {\bibfnamefont {A.}~\bibnamefont {Chambolle}},\ }\bibfield  {title} {\bibinfo {title} {An algorithm for total variation minimization and applications},\ }\href@noop {} {\bibfield  {journal} {\bibinfo  {journal} {Journal of Mathematical imaging and vision}\ }\textbf {\bibinfo {volume} {20}},\ \bibinfo {pages} {89} (\bibinfo {year} {2004})}\BibitemShut {NoStop}%
\bibitem [{\citenamefont {Chen}\ \emph {et~al.}(2023)\citenamefont {Chen}, \citenamefont {Wang},\ and\ \citenamefont {Heidrich}}]{chen2023h}%
  \BibitemOpen
  \bibfield  {author} {\bibinfo {author} {\bibfnamefont {N.}~\bibnamefont {Chen}}, \bibinfo {author} {\bibfnamefont {C.}~\bibnamefont {Wang}},\ and\ \bibinfo {author} {\bibfnamefont {W.}~\bibnamefont {Heidrich}},\ }\bibfield  {title} {\bibinfo {title} {{$\partial \mathbf{H}$}: Differentiable holography},\ }\href@noop {} {\bibfield  {journal} {\bibinfo  {journal} {Laser \& Photonics Reviews}\ }\textbf {\bibinfo {volume} {17}},\ \bibinfo {pages} {2200828} (\bibinfo {year} {2023})}\BibitemShut {NoStop}%
\bibitem [{\citenamefont {Chen}\ \emph {et~al.}(2025)\citenamefont {Chen}, \citenamefont {Brady},\ and\ \citenamefont {Lam}}]{chen2025differentiable}%
  \BibitemOpen
  \bibfield  {author} {\bibinfo {author} {\bibfnamefont {N.}~\bibnamefont {Chen}}, \bibinfo {author} {\bibfnamefont {D.~J.}\ \bibnamefont {Brady}},\ and\ \bibinfo {author} {\bibfnamefont {E.~Y.}\ \bibnamefont {Lam}},\ }\bibfield  {title} {\bibinfo {title} {Differentiable imaging: Progress, challenges, and outlook},\ }\href@noop {} {\bibfield  {journal} {\bibinfo  {journal} {Advanced Devices \& Instrumentation}\ } (\bibinfo {year} {2025})}\BibitemShut {NoStop}%
\bibitem [{\citenamefont {Robinson}\ and\ \citenamefont {Harder}(2009)}]{robinson2009coherent}%
  \BibitemOpen
  \bibfield  {author} {\bibinfo {author} {\bibfnamefont {I.}~\bibnamefont {Robinson}}\ and\ \bibinfo {author} {\bibfnamefont {R.}~\bibnamefont {Harder}},\ }\bibfield  {title} {\bibinfo {title} {Coherent x-ray diffraction imaging of strain at the nanoscale},\ }\href@noop {} {\bibfield  {journal} {\bibinfo  {journal} {Nature Materials}\ }\textbf {\bibinfo {volume} {8}},\ \bibinfo {pages} {291} (\bibinfo {year} {2009})}\BibitemShut {NoStop}%
\bibitem [{\citenamefont {Miao}\ \emph {et~al.}(2011)\citenamefont {Miao}, \citenamefont {Sandberg},\ and\ \citenamefont {Song}}]{miao2011coherent}%
  \BibitemOpen
  \bibfield  {author} {\bibinfo {author} {\bibfnamefont {J.}~\bibnamefont {Miao}}, \bibinfo {author} {\bibfnamefont {R.~L.}\ \bibnamefont {Sandberg}},\ and\ \bibinfo {author} {\bibfnamefont {C.}~\bibnamefont {Song}},\ }\bibfield  {title} {\bibinfo {title} {Coherent x-ray diffraction imaging},\ }\href@noop {} {\bibfield  {journal} {\bibinfo  {journal} {IEEE Journal of Selected Topics in Qantum Electronics}\ }\textbf {\bibinfo {volume} {18}},\ \bibinfo {pages} {399} (\bibinfo {year} {2011})}\BibitemShut {NoStop}%
\bibitem [{\citenamefont {Zheng}\ \emph {et~al.}(2013)\citenamefont {Zheng}, \citenamefont {Horstmeyer},\ and\ \citenamefont {Yang}}]{zheng2013wide}%
  \BibitemOpen
  \bibfield  {author} {\bibinfo {author} {\bibfnamefont {G.}~\bibnamefont {Zheng}}, \bibinfo {author} {\bibfnamefont {R.}~\bibnamefont {Horstmeyer}},\ and\ \bibinfo {author} {\bibfnamefont {C.}~\bibnamefont {Yang}},\ }\bibfield  {title} {\bibinfo {title} {Wide-field, high-resolution {F}ourier ptychographic microscopy},\ }\href@noop {} {\bibfield  {journal} {\bibinfo  {journal} {Nature Photonics}\ }\textbf {\bibinfo {volume} {7}},\ \bibinfo {pages} {739} (\bibinfo {year} {2013})}\BibitemShut {NoStop}%
\bibitem [{\citenamefont {Horstmeyer}\ \emph {et~al.}(2016)\citenamefont {Horstmeyer}, \citenamefont {Chung}, \citenamefont {Ou}, \citenamefont {Zheng},\ and\ \citenamefont {Yang}}]{horstmeyer2016diffraction}%
  \BibitemOpen
  \bibfield  {author} {\bibinfo {author} {\bibfnamefont {R.}~\bibnamefont {Horstmeyer}}, \bibinfo {author} {\bibfnamefont {J.}~\bibnamefont {Chung}}, \bibinfo {author} {\bibfnamefont {X.}~\bibnamefont {Ou}}, \bibinfo {author} {\bibfnamefont {G.}~\bibnamefont {Zheng}},\ and\ \bibinfo {author} {\bibfnamefont {C.}~\bibnamefont {Yang}},\ }\bibfield  {title} {\bibinfo {title} {Diffraction tomography with {F}ourier ptychography},\ }\href@noop {} {\bibfield  {journal} {\bibinfo  {journal} {Optica}\ }\textbf {\bibinfo {volume} {3}},\ \bibinfo {pages} {827} (\bibinfo {year} {2016})}\BibitemShut {NoStop}%
\bibitem [{\citenamefont {Zhou}\ \emph {et~al.}(2022)\citenamefont {Zhou}, \citenamefont {Li}, \citenamefont {Sun}, \citenamefont {Zhou}, \citenamefont {Ullah}, \citenamefont {Bai}, \citenamefont {Chen},\ and\ \citenamefont {Zuo}}]{zhou2022transport}%
  \BibitemOpen
  \bibfield  {author} {\bibinfo {author} {\bibfnamefont {S.}~\bibnamefont {Zhou}}, \bibinfo {author} {\bibfnamefont {J.}~\bibnamefont {Li}}, \bibinfo {author} {\bibfnamefont {J.}~\bibnamefont {Sun}}, \bibinfo {author} {\bibfnamefont {N.}~\bibnamefont {Zhou}}, \bibinfo {author} {\bibfnamefont {H.}~\bibnamefont {Ullah}}, \bibinfo {author} {\bibfnamefont {Z.}~\bibnamefont {Bai}}, \bibinfo {author} {\bibfnamefont {Q.}~\bibnamefont {Chen}},\ and\ \bibinfo {author} {\bibfnamefont {C.}~\bibnamefont {Zuo}},\ }\bibfield  {title} {\bibinfo {title} {Transport-of-intensity {F}ourier ptychographic diffraction tomography: defying the matched illumination condition},\ }\href@noop {} {\bibfield  {journal} {\bibinfo  {journal} {Optica}\ }\textbf {\bibinfo {volume} {9}},\ \bibinfo {pages} {1362} (\bibinfo {year} {2022})}\BibitemShut {NoStop}%
\bibitem [{\citenamefont {Wu}\ \emph {et~al.}(2024)\citenamefont {Wu}, \citenamefont {Zhou}, \citenamefont {Chen}, \citenamefont {Sun}, \citenamefont {Lu}, \citenamefont {Chen},\ and\ \citenamefont {Zuo}}]{wu2024lens}%
  \BibitemOpen
  \bibfield  {author} {\bibinfo {author} {\bibfnamefont {X.}~\bibnamefont {Wu}}, \bibinfo {author} {\bibfnamefont {N.}~\bibnamefont {Zhou}}, \bibinfo {author} {\bibfnamefont {Y.}~\bibnamefont {Chen}}, \bibinfo {author} {\bibfnamefont {J.}~\bibnamefont {Sun}}, \bibinfo {author} {\bibfnamefont {L.}~\bibnamefont {Lu}}, \bibinfo {author} {\bibfnamefont {Q.}~\bibnamefont {Chen}},\ and\ \bibinfo {author} {\bibfnamefont {C.}~\bibnamefont {Zuo}},\ }\bibfield  {title} {\bibinfo {title} {Lens-free on-chip {3D} microscopy based on wavelength-scanning {F}ourier ptychographic diffraction tomography},\ }\href@noop {} {\bibfield  {journal} {\bibinfo  {journal} {Light: Science \& Applications}\ }\textbf {\bibinfo {volume} {13}},\ \bibinfo {pages} {237} (\bibinfo {year} {2024})}\BibitemShut {NoStop}%
\bibitem [{\citenamefont {Hololab}(2025)}]{GS_github}%
  \BibitemOpen
  \bibfield  {author} {\bibinfo {author} {\bibfnamefont {T.}~\bibnamefont {Hololab}},\ }\href {https://github.com/THUHoloLab/Gaussian-Splats-Holography} {\bibinfo {title} {{Gaussian Splatting Holography}}} (\bibinfo {year} {2025})\BibitemShut {NoStop}%
\end{thebibliography}%

\end{document}